\documentclass[11pt,twoside]{article}

\usepackage{asp2006}
\usepackage{epsf}
\usepackage{epsfig}
\usepackage{lscape}
\begin{document}
\title{A Dynamical Miss:  A Study of the Discrepancy Between Optical and Infrared Kinematics in Mergers}   
\author{Barry Rothberg}   
\affil{Naval Research Laboratory, Code 7211, 4555 Overlook Ave SW, Washington D.C. 20375}    

\begin{abstract} 
Recently, controversy has erupted over whether gas-rich spiral-spiral mergers are capable of forming 
{\it m$^{*}$} ellipticals. Measurements of $\sigma$$_{\circ}$ from the 
2.29$\micron$ CO band-head for local LIRG/ULIRGs, suggest they are not.  IR-bright mergers are often cited as the best candidates 
for forming massive ellipticals, so the recent observations have raised doubts about both the Toomre Merger Hypothesis and the 
fundamental assumptions of $\Lambda$-CDM galaxy formation models. However, kinematics obtained with the 
Calcium II Triplet at 8500 {\AA} suggest mergers are forming {\it m} $\ge$ {\it m$^{*}$} ellipticals.  In this work, 
we show that kinematics derived from the CO stellar absorption band-head leads to a significant underestimation of the masses of LIRGs/ULIRGs.  
This is primarily due to the presence of a young population affecting CO band-head measurements.
\end{abstract}

\section{Introduction}  
\indent In the local universe, Luminous and Ultra-luminous Infrared Galaxies (LIRGs \& ULIRGs) 
have long been proposed as ideal candidates for forming massive 
elliptical galaxies \citep{1992ApJ...390L..53K}. They contain vast quantities
of molecular gas and most show evidence of recent or 
ongoing merging activity, along with relatively
high star-formation rates. Observations of these and other 
advanced mergers have long suggested that their luminosities are equivalent to
or greater than massive elliptical galaxies 
\citep{1982ApJ...252..455S,1996AJ....111..109S,1996AJ....111..655H}, and \cite{2004AJ....128.2098R},
hereafter Paper I.\\
\indent The arguments {\it against} LIRGs/ULIRGs, and mergers in general forming {\it L$^{*}$} 
or {\it m$^{*}$} elliptical galaxies are based on kinematic arguements.  
Until recently, kinematic studies of LIRGs/ULIRGs have used the
the infrared CO stellar absorption lines at 1.6$\micron$ and 2.29$\micron$ to measure
central velocity dispersions ($\sigma$$_{\circ}$). These stellar band-heads are prominent in 
late-type stars and lie within observable infrared atmospheric windows.\\
\indent $\sigma$$_{\circ, \rm CO}$'s have been previously measured for $\sim$ 60 LIRGs/ULIRGs and two
non-LIRG/ULIRG merger remnants \citep{1994ApJ...437L..23D,1994ApJ...433L...9S,
1998ApJ...497..163S,1999MNRAS.309..585J,2001ApJ...563..527G,2002ApJ...580...73T,
2006ApJ...651..835D}, yielding a median $\sigma$$_{\circ}$ $\sim$ 150 km s$^{-1}$, far less than an {\it m$^{*}$} elliptical galaxy.\\
\indent \cite{1986ApJ...310..605L} used the Mg{\it Ib} ($\lambda$ $\sim$ 5200 {\AA}) 
and \ion{Ca}{II} triplet ($\lambda$ $\sim$ 8500 {\AA}, hereafter CaT) 
to measure $\sigma$$_{\circ}$ in a sample of 13 visually selected mergers, producing a median of $\sigma$$_{\circ}$ $=$ 200 km s$^{-1}$. 
\cite{2006AJ....131..185R}, hereafter Paper II, measured $\sigma$$_{\circ}$ for 38 optically selected single-nuclei 
merger remnants, including 10 LIRGs and 2 ULIRGs, using the CaT absorption lines and found
a median $\sigma$$_{\circ}$ $=$ 211 km s$^{-1}$ for the entire sample  and a median of 196 km s$^{-1}$ for the LIRGs/ULIRGs.\\
\indent In Paper II, {\it K}-band photometry from Paper I was combined with $\sigma$$_{\circ, \rm CaT}$ to test 
whether and where merger remnants lie on the {\it K}-band Fundamental Plane \citep{1998AJ....116.1606P}.
Most of the merger remnants did lie on or within the scatter of the Fundamental Plane.  
A small number of predominantly LIRG/ULIRG remnants sat offset from the Fundamental Plane in a tail-like feature,
(see Figure 1 in Paper II).  The offset was due to high surface brightness ($<$$\mu$$_{\rm K}$$>$$_{\rm eff}$),  
not small $\sigma$$_{\circ}$, suggesting the presence of a younger IR-bright population.\\
\indent Paper II also demonstrated that the CO band-head produced consistently {\it smaller}
$\sigma$$_{\circ}$ for LIRGs/ULIRGs than CaT.  This suggested previous infrared studies may have 
{\it underestimated} their masses.  Unfortunately, to date, {\it all} published $\sigma$$_{\circ, \rm CaT}$ and $\sigma$$_{\circ, \rm CO}$ 
of remnants lie within the above-mentioned tail-like offset from the {\it K}-band Fundamental Plane, 
making it difficult to separate whether the discrepancy is related to galaxy properties or problems with 
one of the stellar lines.  \cite{2003AJ....125.2809S} (hereafter SG03) first noted that in their sample of 
25 (mostly S0) early-type galaxies, a signficant fraction also showed $\sigma$$_{\circ, \rm CO}$ $<$ $\sigma$$_{\circ, \rm Optical}$.  
The largest discrepancies were found in S0s. They proposed that dust enshrouded stellar disks, visible only in the IR
produced smaller $\sigma$$_{\circ,CO}$.

\section{Sample Selection and Observations}
\indent An E0 and 6 non-LIRG/ULIRG merger remnants found to lie on the Fundamental Plane 
were observed in queue mode with the Gemini Near-Infrared Spectrograph (GNIRS) on Gemini-South 
(Program GS-2007A-Q-17, P.I.  Rothberg) using  the Short Camera with the 111 l/mm 
grating and 0{$\arcsec$}.3 $\times$ 99{$\arcsec$} slit 
(R $\sim$ 6200). Observations were centered on the $^{12}$CO(2,0) 2.29$\micron$ band-head.
Previously published optical and infrared spectroscopic data were also used (see Papers I and II for details), 
bringing the total to 8 non-LIRG and 6 LIRG merger remnants. All $\sigma$$_{\circ}$ were
either extracted or corrected (for data from the literature) to a 1.53 {\it h}$^{-1}$ kpc central aperture.\\
\indent Low-resolution (R $\sim$ 1200) spectra were obtained with SpeX on the 
NASA Infrared Telescope Facility for 11 merger remnants.  Observations were made using 
the short cross-dispersed mode (0.8$\micron$ $<$ $\lambda$ $<$ 2.4$\micron$, {\it R} $\sim$ 1200) 
with the 0$\arcsec$.5 $\times$ 15$\arcsec$ slit.  These observations were used to measure Brackett $\gamma$ (Br$\gamma$)
equivalent widths (EW) and supplement measurement of CO indices for published $\sigma$$_{\circ, \rm CO}$.\\
\indent A comparison sample of 23 pure elliptical galaxies
was assembled from the literature to look for the same discrepancy between $\sigma$$_{\circ, Optical}$ and $\sigma$$_{\circ, \rm CO}$ 
observed in merger remnants and SG03.  
All $\sigma$$_{\circ}$ obtained from the literature were corrected to a 1.53 {\it h}$^{-1}$ kpc aperture.

\section{Comparison between Optical and Infrared $\sigma$$_{\circ}$}
Figure 1 shows a three-panel comparison among the LIRG merger remnants ({\it left}), non-LIRG 
merger remnants ({\it center}) and elliptical galaxies ({\it right}).    
The data were fit with a double weighted least-squares fit to compare with a slope of unity 
to search for discrepancies in $\sigma$$_{\circ}$. Both the non-LIRG merger remnants and the elliptical galaxies
have a slope within 1$\sigma$ of unity, while the the LIRG merger remnants yield a slope of 0.24$\pm$0.1.
\begin{figure}
\hspace{-0.35in}
\epsfig{file=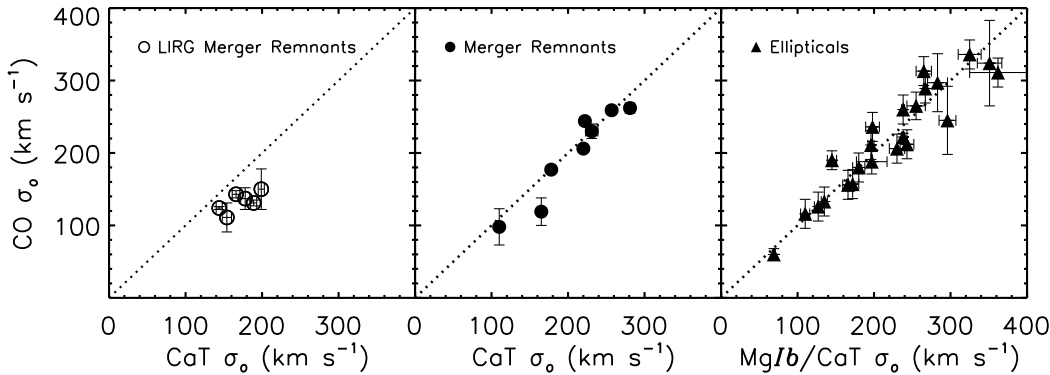,scale=0.7}
\vspace{-0.2in}
\caption{Comparison between $\sigma$$_{\circ, \rm Optical}$ and  $\sigma$$_{\circ, \rm CO }$
for LIRG ({\it left}) and non-LIRG ({\it center}) merger remnants and 
elliptical galaxies ({\it right}).  The overplotted dotted line represents $\sigma$$_{\circ, \rm Optical}$ $=$  $\sigma$$_{\circ, \rm CO }$.}
\end{figure}

\section{$\sigma$$_{\circ}$ Variations Compared with Stellar Populations}
\indent In Paper II it was postulated that the CaT and CO stellar features were probing two different
stellar populations; the former probing older, late-type giants contributed from the progenitor spirals, 
and the latter sensitive to young Red Supergiants (RSGs) or Asymptotic Giant Branch (AGB) stars formed during the merger. 
A direct way to test this is to measure EWs of the CaT and CO features and compare those
with stellar populations models in order to age-date each population.  The premise is that the younger stellar populations 
are created within a disk formed when gas from the progenitor spirals dissipates towards the barycenter of the 
merger (see \cite{2002MNRAS.333..481B}).  At {\it K}-band, the light from this population dominates during
certain epochs \citep{2005MNRAS.362..799M} (hereafter M05).\\
\indent Figure 2 shows stellar population models from M05 ({\it solid lines}) for a simple burst population with a Salpeter IMF and solar metallicity
for the CaT$^{*}$ and CO indices.  CaT$^{*}$ EWs and CO$_{\rm phot}$ indices were measured for each merger remnant  
and used in conjunction with the models to age-date each remnant as plotted.
In some cases, the measured indices intersected multiple points in the models.  
The degeneracy was broken by measuring the EW of the Paschen triplet (PaT) absorption lines (near 8500 {\AA}) and comparing with predicted values 
from M05, or measuring the EWs of Br$\gamma$ emission (2.165$\micron$) and comparing with predicted values from Starburst99 \citep{1999ApJS..123....3L}.
In some cases the degeneracy for CO ages could not be broken due to the absence of Br$\gamma$.  Double values
are plotted in Figure 2 with a dotted line connecting them.\\
\indent Overplotted in Figure 2 ({\it right}) next to each merger remnant are the fractional $\sigma$$_{\circ}$ differences 
($\sigma$$_{\circ, CaT}$ - $\sigma$$_{\circ, CO}$)/$\sigma$$_{\circ, CaT}$.  
The LIRG merger remnants show larger fractional differences than the non-LIRG remnants (except NGC 7252).
The results in Figure 2 appear to support the premise that the presence of young populations
are responsible for the smaller $\sigma$$_{\circ}$ observed in LIRGs. 
The final test will be to confirm whether these young populations reside in rotating disks located within the central few kpc of the LIRG/ULIRG 
merger remnant using spatially resolved spectroscopic observations of the CO band-head.
\begin{figure}
\hspace{-0.5in}
\epsfig{file=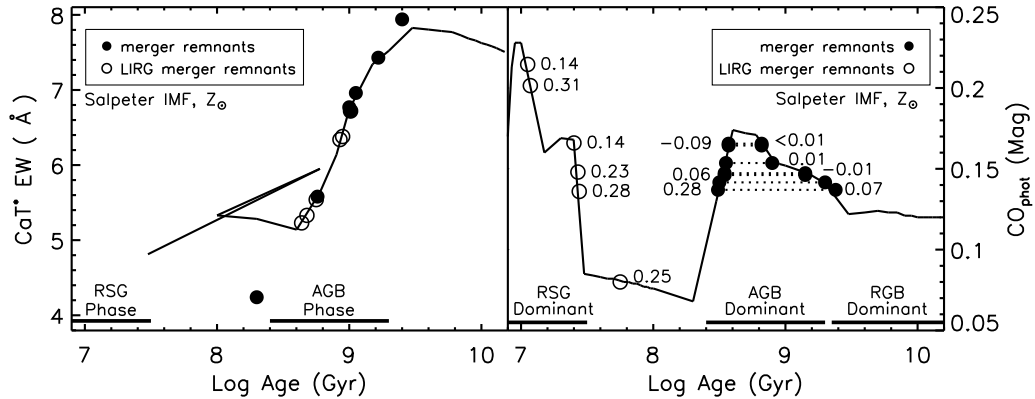,scale=0.7}
\vspace{-0.2in}
\caption{Stellar population models from M05 ({\it solid lines}) for CaT$^{*}$ ({\it left}) and CO ({\it right}) 
indices.  Overplotted are 6 LIRG merger remnants ({\it open circles}) and 7 non-LIRG merger remnants ({\it filled circles}).  
RSG and AGB phases are noted along with when phases dominate the {\it K}-band light.
The numbers in the right panel are ($\sigma$$_{\circ, CaT}$ - $\sigma$$_{\circ, CO}$)/$\sigma$$_{\circ, CaT}$ .  Dotted lines indicate degenerate age results. }
\end{figure}


\end{document}